# Magnetic skyrmion-based synaptic devices

Yangqi Huang[a†], Wang Kang[a†], Xichao Zhang[b], Yan Zhou[b] and Weisheng Zhao[a*]

**Magnetic skyrmions are promising candidates for next-generation information carriers, owing to their small size, topological stability, and ultralow depinning current density. A wide variety of skyrmionic device concepts and prototypes have been proposed, highlighting their potential applications. Here, we report on a bio-inspired skyrmionic device with synaptic plasticity. The synaptic weight of the proposed device can be strengthened/weakened by positive/negative stimuli, mimicking the potentiation/depression process of a biological synapse. Both short-term plasticity (STP) and long-term potentiation (LTP) functionalities have been demonstrated for a spiking time-dependent plasticity (STDP) scheme. This proposal suggests new possibilities for synaptic devices for use in spiking neuromorphic computing applications.**

Neuromorphic computing, inspired by the biological nervous system, has attracted considerable attention recently[1-5]. Owing to its massively paralleled nature, power efficiency, robustness against fault/variation, and combination of memory and computation, it is highly efficient in cognition- and perception-related tasks[1]. One of the key elements of the human brain is the synapse, which connects two neurons with plastic strength (or weight)[1]. In recent decades, a variety of nanoelectronic devices have been developed to emulate synapses, including phase-change memories[6,7], Ag-Si memories[8], and resistive memories[9–11], etc. Recently, some devices with features analogous to biological synapses have emerged, offering new opportunities for artificial synapse design.

Magnetic skyrmions are nanoscale particle-like spin textures that are topologically stable. Nanoscale skyrmions are typically stabilized by the Dzyaloshinskii–Moriya interaction (DMI) in non-centrosymmetric bulk magnets or magnetic thin films[12–14]. Since the first experimental observation of skyrmions in 2009[15], they have attracted extensive research interest, from the perspectives of physical fundamentals to electronic applications. Owing to their small size, robustness against defects, and low depinning current density[16,17], skyrmions have considerable potential for use as information carriers in future ultra-dense and low-power spintronic devices, such as racetrack memories[18,19] and logic gates[20,21]. Amazing theoretical and experimental advances have been made in skyrmion research, and a wide variety of skyrmionic device concepts and prototypes have been proposed[18–22], reflecting the considerable potential of skyrmions for future spintronic applications. Recently, stable skyrmions have been experimentally observed and manipulated at room temperature[23–28], further highlighting their potential.

One of the intrinsic features of skyrmions is their particle-like behavior, due to which multiple skyrmions can aggregate, exhibiting potential for multi-valued storage devices with skyrmions as information carriers. The states of such devices can be modulated by an electric current that drives skyrmions into or out of the devices. Such characteristics are analogous to those of biological synapses, in that their weight can be dynamically modulated based on the temporal correlation (or stimulus) between the spiking activities of the interconnecting pre- and post-neurons, i.e., the synaptic plasticity. This behavior is believed to be the basis for learning and memory in the human brain[29–31]. In a biological synapse, synaptic plasticity is achieved based on signal propagation by the release of neurotransmitters[32]. For artificial synapses, the capability responsible for short-term plasticity (STP), long-term potentiation (LTP), and temporal dynamics is the foundation of learning in spiking neuromorphic applications.

In this communication, we propose a novel bio-inspired skyrmionic device with synaptic plasticity. The weight of the proposed skyrmionic device can be modulated by an electric current and measured through the magnetoresistance effect. The polymorphism, synaptic plasticity (both STP and LTP), and spike timing-dependent plasticity (STDP) of the proposed device were investigated through micromagnetic simulations.

## Skyrmionic Synapse Structure

Figure 1(b) shows a schematic of the proposed skyrmionic synaptic device. The primary components of the device are a ferromagnetic (FM) layer on a heavy metal (HM) and a gated

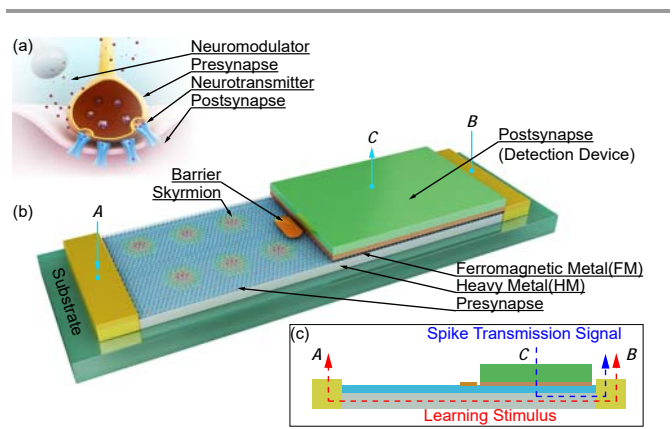

*Fig. 1* Schematic of (a) a biological synapse and (b) the proposed skyrmionic synaptic device. To mimic a neuromodulator, a bidirectional learning stimulus flowing through the HM from terminal A to terminal B (or vice versa) drives skyrmions into (or out of) the postsynapse region to increase (or decrease) the synaptic weight, as shown in (c), mimicking the potentiation/depression process of a biological synapse[3]. A detection device at terminal C measures the weight of the synaptic device via the magnetoresistance effect, which also modulates the post-synaptic spike current (depending on the weight of the synaptic device).

[a.] *Fert Beijing Institute, BDBC, and School of Electronic and Information Engineering, Beihang University, Beijing, China. E-mail: wang.kang@buaa.edu.cn, and weisheng.zhao@buaa.edu.cn*

[b.] *School of Science and Engineering, Chinese University of Hong Kong, Shenzhen, China. E-mail: yanzhou@hku.hk*

† *Yangqi Huang and Wang Kang contributed equally to this work and therefore can be considered co-first authors.*

barrier. The FM layer has perpendicular magnetic anisotropy (PMA), and the DMI is generated at the interface between the FM layer and the HM. The FM layer and the HM together form a nanotrack for skyrmion motion. The gated barrier, which is located at the center of the nanotrack, is designed to have a higher PMA than the FM layer, separating the entire nanotrack into two regions, referred to as the presynapse and postsynapse regions, on the left-hand and right-hand sides of the barrier, respectively. This device structure is analogous to a biological synapse, as shown in Fig. 1(a). To illustrate the synaptic plasticity of the device, a side view of the device is shown in Fig. 1(c). In the spike transmission mode of operation, the spike produced by a pre-neuron is modulated by the weight (magnetoresistance) of the synaptic device, which generates a post-synaptic spike current from terminal C to terminal B. During the learning phase, a bidirectional charge current flows through the HM, injecting a vertical spin current into the FM layer from terminal A to terminal B (or vice versa) and driving skyrmions into (or out of) the postsynapse region to increase (or decrease) the synaptic weight, mimicking the potentiation/depression process of a biological synapse. The proposed device provides decoupled spike-transmission and learning channels (see Fig. 1(c)). The channel between terminal A and terminal B, which is called a learning channel, is used to transmit learning stimuli from the pre-neurons. The channel between terminal C and terminal B, which is called the spike transmission channel, is used to transmit a spike current to the post-neurons. A detection device is located on top of the postsynapse region to read the weight (magnetoresistance) of the synapse through the magnetoresistance effect[33].

Figure 2 illustrates the results of micromagnetic simulations of the three primary operation modes of our proposed synaptic device: the initialization mode, the potentiation mode, and the depression mode. Before we illustrate the operation modes of the proposed synaptic device, we define two terms: (a) positive stimulus, which signifies an electric current with an amplitude of 5 $MA/cm^2$ flowing from terminal A to terminal B; and (b) negative stimulus, which signifies an electrical current with amplitude of 5 $MA/cm^2$ flowing from terminal B to terminal A. In the initialization mode (from 0 to 35 ns), skyrmions are generated in the presynapse region of the device. Owing to the repulsion between skyrmions and the nanotrack edges[34], a threshold value for the total number of skyrmions (11 skyrmions in our design, with a 120-nm-wide nanotrack) in the presynapse region of the device will be reached. This threshold value determines the programming resolution of the synaptic weight of the device. It is worth noting that the initialization is performed only once. Because skyrmions are non-volatile spin textures, they will be maintained even if the power is off.

In the potentiation mode (from 35 to 65 ns; see Fig. 2(b)), a positive stimulus drives skyrmions from the presynapse region into the postsynapse region, increasing the synaptic weight of the device. In a similar way, during the depression mode (from 87 to 117 ns; see Fig. 2(c)), a negative stimulus drives skyrmions from the postsynapse region into the presynapse region, decreasing the synaptic weight of the device. The red curve in Fig. 2(d) shows the normalized mz (the average magnetization component in the z direction) of the postsynapse region of the device. The shifting of the mz corresponds to the variation in the skyrmion number and size. It should be noted that two skyrmions fail to pass through the barrier in both the potentiation and depression modes. This can be explained as a consequence of the insufficiency of the total driving force, which consists of the driving force of the electric current and the repulsion force of the skyrmions. Taking the potentiation mode as an example, as skyrmions move into the postsynapse region, the repulsion force of the skyrmions in the presynapse region, which favors skyrmion motion from the presynapse region into the postsynapse region, will decrease, whereas the repulsion force of the skyrmions in the postsynapse region, which hinders skyrmion motion from the presynapse region into the postsynapse region, will increase. Meanwhile, the driving force of the electric current (herein we consider a direct current, DC), which depends on the current density, and the repulsion force of the barrier remain unchanged. Finally, all these forces enter an equilibrium state, leaving two skyrmions in the presynapse region. The synaptic weight of the postsynapse region of the device can be determined by measuring the magnetoresistance through the detection device at terminal C[22,35,36]. This synaptic weight change behavior, illustrated in Fig. 2(d), demonstrates the stimulus-induced synaptic plasticity of the proposed skyrmionic device. It is worth noting that the size of the skyrmions extruded depends on the driving force and the repulsion force, as the snapshots in Fig. 2 show. As soon as the stimulus is turned off, the compressed skyrmions began to expand into an equilibrium state, leading to an obvious change in the mz of the post synapse region.

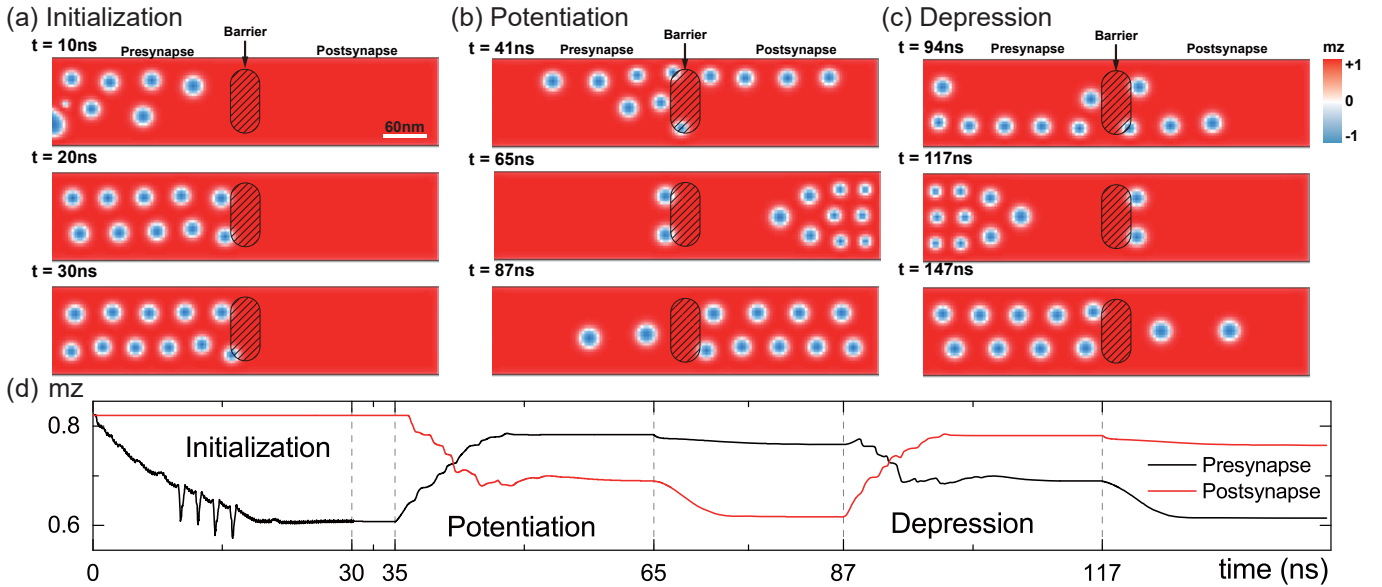

*Fig. 2* Micromagnetic simulations of the operation modes of our proposed skyrmionic synaptic device. (a) Initialization (from 0 to 35 ns): skyrmions are generated in the presynapse region; (b) potentiation mode (from 35 to 87 ns): a positive stimulus (from 35 to 65 ns) drives skrymions from the presynapse region to the postsynapse region, increasing the synaptic weight of the device; and (c) depression mode (from 87 to 147 ns): a negative stimulus (from 87 to 117 ns) drives skrymions from the postsynapse region into the presynapse region, decreasing the synaptic weight of the device. In each operation mode, the device is relaxed to the equilibrium state. (a), (b), and (c) show snapshots of the magnetization of the nanotrack, and (d) shows the time-resolved normalized mz (the average magnetization component in the z direction) of the presynapse and postsynapse regions.

Figure 3(a) shows the dependence of the saturation number of skyrmions during the initialization mode with respect to the size of the presynapse region of the device. The length of the presynapse region was fixed. Because the area of the postsynapse region increases linearly, a linear model was used to characterize the relation between the nanotrack width and the number of saturation skyrmions. Intuitively, a wider nanotrack can accommodate more skyrmions in the device, indicating a larger capability of synaptic weight resolution. However, there are trade-offs among the synaptic weight resolution, processing speed, area, and power consumption. The reason for this is that a wider nanotrack (with more skyrmions) requires a longer time than a narrower nanotrack to achieve a saturation state during the initialization mode and to change the same percentage of synaptic weight during the learning operation. In addition, a wider nanotrack requires a larger current amplitude to achieve the same current density, consuming more energy. Figure 3(b) shows the relationship between the programming speed (i.e., the number of skyrmions in the postsynapse region of the device) and the driving current density. Obviously, a larger current density results in a fast skyrmion motion and thus a higher programming speed.

We also investigated in detail the effects of the synaptic plasticity of the proposed device on the stimulus pulse characteristics. We considered three stimulus configurations: case 1, 1.5 ns in duration at 5-ns intervals, as shown in Fig. 4(b); case 2, 1 ns in duration at 2-ns intervals, as shown in Fig. 4(c); and case 3, 1 ns in duration at 5-ns intervals, as shown in Fig. 4(d). Every configuration consists of eight stimulus pulses. We monitored the magnetoresistance change of the postsynapse region of the device. As Fig. 4 shows, cases 1 and 2 demonstrates an LTP property, while case 3 demonstrates an STP property[5,37]. Compared with cases 1 and 3, with the same interval, a proper stimulus duration is required to transfer STP to LTP. Cases 2 and 3 demonstrate that the interval of the stimulus also plays an essential role in the device's plasticity.

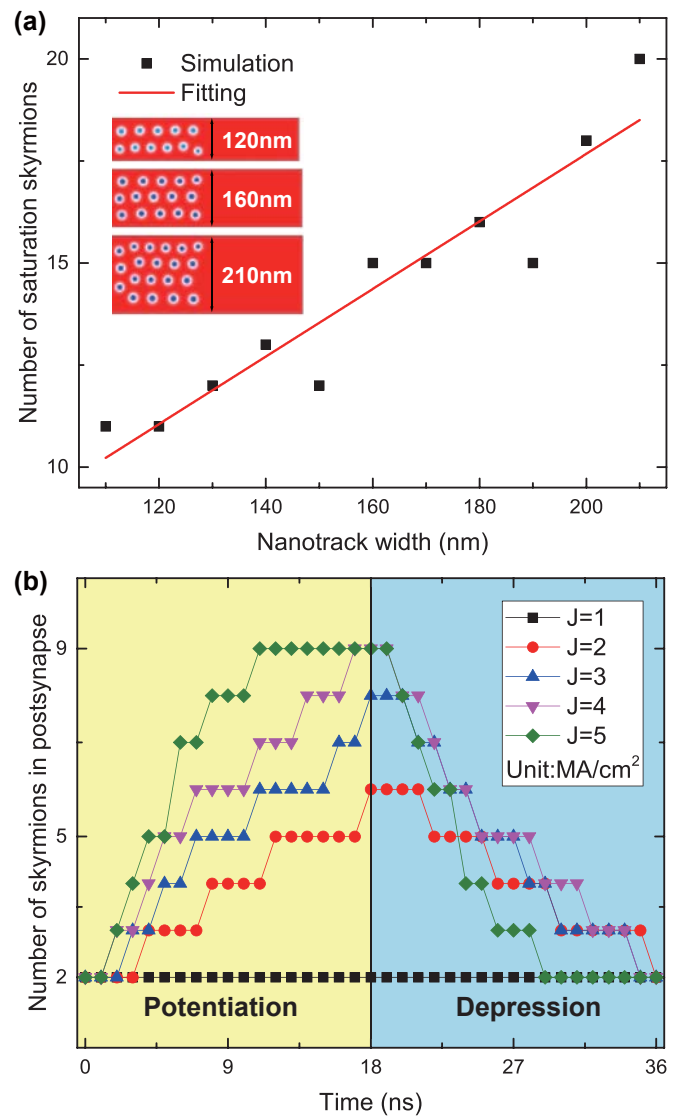

*Fig. 3* (a) Saturation number of skyrmions after initialization with respect to the nanotrack width. (b) Number of skyrmions in the postsynapse during a potentiation mode followed by a depression mode. Different curves correspond to different learning stimuli densities.

The STP and LTP of the proposed synaptic device can be explained by the competition between the driving force provided by the electric current and the repulsion force provided by the barrier when the skyrmion passes through the barrier. When the skyrmion approaches the barrier, the repulsion force of the barrier increases, leading to an energy barrier, as shown in the energy profile of Fig. 4(a). Upon receipt of an input stimulus, the driving force of the electric current proceeds uphill along the force energy profile (from point 1 to point 2). However, because point 2 is a meta-stable state, it falls to point 1 once the stimulus is removed. If the input stimulus is not of sufficient duration and frequency, the skyrmion will try to move back to the initial position after each stimulus, corresponding to STP. Otherwise, if the stimulus is of sufficient duration and/or frequency, the skyrmion will not have enough time to reach point 1 and ultimately will be able to pass the barrier to point 3, corresponding to LTP. Once the skyrmion passes through the barrier, it is difficult for the skyrmion to exhibit STP and move back to the initial position. This behavior is consistent with the psychological model of a biological synapse.

## Simulation and discussion

The proposed skyrmionic synaptic device was numerically simulated by solving the Landau–Lifshitz–Gilbert (LLG) equation with spin transfer torques as follows[14]:

$$\frac{d\bm{m}}{dt} = -|\gamma|\bm{m} \times \bm{h}_{eff} + \alpha \bm{m} \times \frac{d\bm{m}}{dt} + \frac{u}{t}\bm{m} \times (\bm{m}_p \times \bm{m})$$

where $\bm{m} = \bm{M}/M_s$ is the reduced magnetization, $M_s = 580 kA/m$ is the saturation magnetization, $\gamma = -2.211 \times 10^5 mA^{-1}s^{-1}$ is the gyromagnetic ratio, $\bm{h}_{eff} = \bm{H}_{eff}/M_s$ is the reduced effective field, $\alpha = 0.3$ is the Gilbert damping, t is the thickness of the FM layer, $u = \gamma(\hbar jP/2eM_s)$, j is the density of the spin current, and $P = 0.4$ is the spin polarization. The parameters are adopted from Ref. 3.

The nanotrack was designed with a default size of $528\ nm \times 120\ nm \times 1\ nm$, an exchange stiffness $A = 15\ pJ/m$, a PMA constant $K_u = 0.7\ MJ/m^3$, and a DMI constant of $D = 3$. The PMA of the gated barrier was set to be 1.2 times that of the FM. The barrier was a rounded rectangle with dimensions of $40\ nm \times 56\ nm$ and a higher PMA constant of $K_u = 0.84\ MJ/m^3$. A discretization of $2\ nm \times 2\ nm \times 1\ nm$ was used in our simulation.

As information carriers, skyrmions are naturally non-volatile and particle-like. Similar to the activation of N-methyl-D-aspartate (NMDA) receptors in the hippocampus, which is the foundation of synaptic plasticity[30], transmitting a skyrmion from the presynapse region to the postsynapse region leads to both STP and LTP functionalities. The saturation number of skyrmions and the size of skyrmions in the equilibrium state can be designed using different materials and nanotrack widths to achieve different resolutions of the synaptic weight in our devices.

## Conclusions

We have proposed a bio-inspired skyrmionic device with

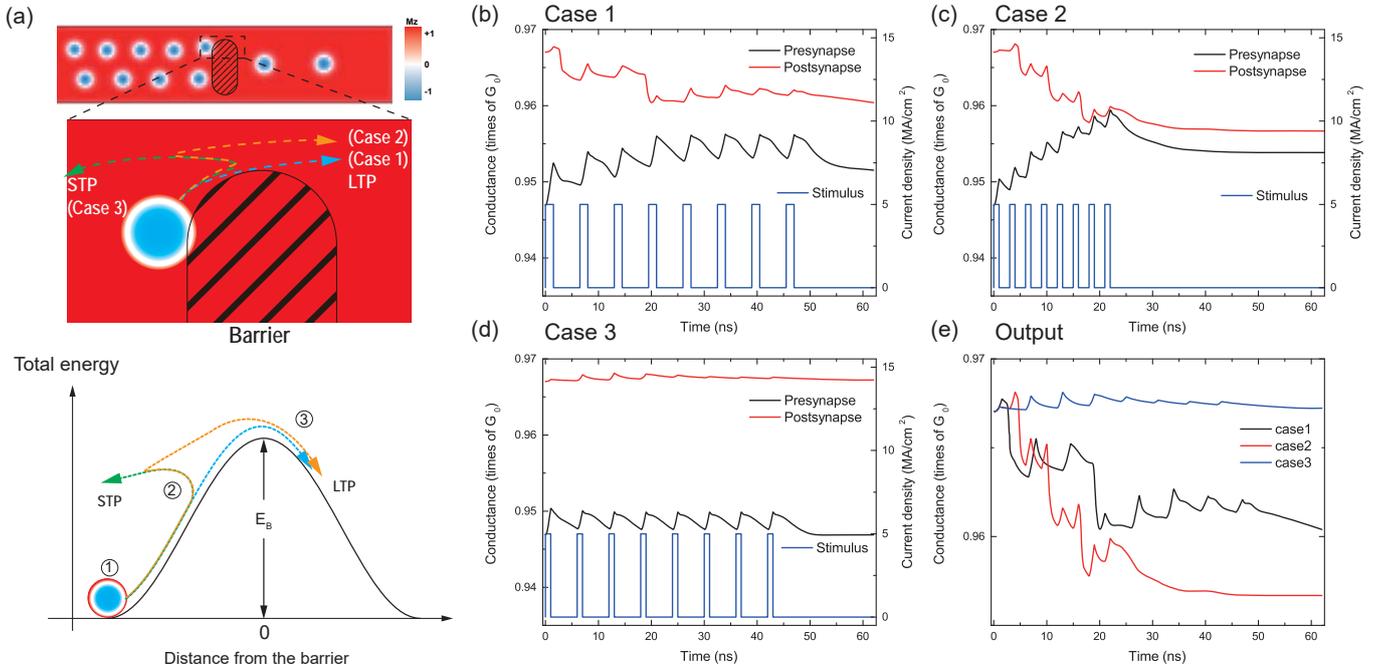

*Fig. 4* (a) Illustration of STP and LTP of the proposed synaptic device. (b), (c), and (d) Conductance variations of the postsynapse under different stimulus pulses: (b) case 1, 1.5 ns in duration with 5-ns interval; (c) case 2, 1 ns in duration with 2-ns interval; and (d) case 3, 1 ns in duration with 5-ns interval. (e) Comparison of the conductance change rate among the three cases.

synaptic plasticity for spiking neuromorphic applications. The spike-transmission and learning operations of the proposed device were demonstrated with micromagnetic simulations. The STP, LTP, and STDP functions of the device were illustrated and discussed. The resolution of the synaptic weight can be adjusted based on the nanotrack width and the skyrmion size. The proposed device suggests new possibilities for the use of skyrmionic devices in neuromorphic applications.

## Notes and references